\documentclass[journal]{IEEEtran}
\IEEEoverridecommandlockouts                              % This command is only
\usepackage[dvipsnames]{xcolor}

\usepackage{algorithm}
\usepackage{algorithmic}
\usepackage[algo2e]{algorithm2e} 

\usepackage{times}
\usepackage{graphicx}
\usepackage{amssymb}
\usepackage{gensymb}
\usepackage{amsmath}
\usepackage{url,hyperref}

\usepackage[labelfont={bf},font=small]{caption}
\usepackage[none]{hyphenat}

\usepackage{mathtools, cuted}

\usepackage[noadjust, nobreak]{cite}
\usepackage{tabularx}
\usepackage{amsmath}

\usepackage{float}

\usepackage{pifont}% http://ctan.org/pkg/pifont

\newcolumntype{Y}{>{\centering\arraybackslash}X}

\usepackage[]{placeins}

\usepackage{placeins}

\usepackage{tikz}

\usepackage[framemethod=tikz]{mdframed}

\usepackage{afterpage}

\usepackage{stfloats}

\usepackage{atbegshi}
\newcommand{\handlethispage}{}
\newcommand{\discardpagesfromhere}{\let\handlethispage\AtBeginShipoutDiscard}
\newcommand{\keeppagesfromhere}{\let\handlethispage\relax}
\AtBeginShipout{\handlethispage}

\usepackage{comment}

\title{\huge{Database Assisted Nonlinear Least Squares Algorithm for Visible Light Positioning in NLOS Environments}}

\author{Ahmet Faruk Saz, \textit{Student Member, IEEE}, and Sinan\thanks{A. F. Saz and S. Gezici are with the  Department of Electrical and Electronics Engineering,  Bilkent  University, 06800,  Ankara, Turkey  (e-mails: faruk.saz@ug.bilkent.edu.tr, gezici@ee.bilkent.edu.tr).} Gezici, \textit{Senior Member, IEEE}}

\usepackage{textgreek}
\begin{document}

\maketitle
\thispagestyle{empty}
\pagestyle{empty}

%%%%%%%%%%%%%%%%%%%%%%%%%%%%%%%%%%%%%%%%%%%%%%%%%%%%%%%%%%%%%%%%%%%%%%%%%%%%%%%%
\begin{abstract}
We propose an indoor localization algorithm for visible light systems by considering effects of non-line-of-sight (NLOS) propagation. The proposed algorithm, named database assisted nonlinear least squares (DA-NLS), utilizes ideas from both the classical NLS algorithm and the fingerprinting algorithm to achieve accurate and robust localization performance in NLOS environments. In particular, a database is used to learn NLOS  effects, and then an NLS algorithm is employed to estimate the position. The performance of the proposed algorithm is compared against that of the fingerprinting and NLS algorithms.

\end{abstract}

%%%%%%%%%%%%%%%%%%%%%%%%%%%%%%%%%%%%%%%%%%%%%%%%%%%%%%%%%%%%%%%%%%%%%%%%%%%%%%%%

\section{Introduction}\label{sec:Intro}

In visible light positioning (VLP) systems, optical signals transmitted from light emitting diodes (LEDs) are received by photo detectors (PDs) or imaging sensors, and are processed to extract location information. VLP systems can facilitate accurate localization and provide low-cost solutions for various location aware applications \cite{sinan4,SurveyTut}.

The position estimation technique commonly employed in VLP systems is the two-step positioning approach, in which position related parameters are extracted from received signals in the first step, and position estimation based on those parameters is performed in the second step \cite{sinanDirect}. Among various position related parameters are received signal strength (RSS), time of arrival (TOA), time difference of arrival (TDOA), and angle of arrival (AOA). In VLP systems, the RSS parameter (equivalently, received power) is commonly employed since it can be estimated in a low complexity manner (as synchronization is not necessary) and carries accurate position related information \cite{sinanDirect,sinan5}. In the second step of the two-step positioning approach, statistical and fingerprinting (mapping) techniques can be used for position estimation. Statistical techniques utilize the statistical properties of the parameters obtained in the first step to design position estimators based on maximum likelihood or Bayesian estimation principles \cite{sinanDirect,Guvenc_hybrid}. For example, when the parameters are corrupted by independent and zero mean Gaussian noise components, the maximum likelihood approach leads to the well-known nonlinear least-squares (NLS) estimator \cite{fund_limits_ref8}. The NLS estimator has favorable performance in line-of-sight (LOS) environments as the zero mean Gaussian noise model is well-suited in such scenarios \cite{fund_limits}. Also, several modifications can be applied to the NLS estimator so as to improve its performance in non-line-of-sight (NLOS) environments \cite{fund_limits_ref8,SurveyNLOS}.
{Consideration of NLOS propagation can be crucial for accurate localization \cite{soft1,soft2}.} On the other hand, in the fingerprinting approach, a database is formed in the offline (training) phase based on parameter estimates at known locations, and then  parameter estimates obtained in the online phase are used together with the database to perform position estimation via a learning algorithm such as $k$-nearest neighbor ($k$-NN), support vector regression (SVR), or neural networks \cite{Fingerprint_2012,Hosseinianfar,Amorph_2016,kNN_compare,FingerReg,MLVLP,LSSVM}. For example, \cite{Hosseinianfar} forms a database of the first and second peaks of each received signal and the time delay between them, and utilizes this database for position estimation in the online phase via the $1$-NN algorithm.

%of channel impulse responses (corresponding to LOS and first reflection of the NLOS powers) and time delay between the responses.

%In order to improve the accuracy of fingerprinting algorithm, several modifications such as utilizing channel state information (CSI) \cite{fing1}, improving database accuracy via crowdsourcing \cite{fing3,fing4} or combining triangulation and fingerprinting \cite{fing2,fing4} are proposed in previous works.

In this letter, we propose a localization algorithm for VLP systems by combining the ideas in the NLS and fingerprinting algorithms in order to achieve accurate and robust localization performance in NLOS environments. In the proposed algorithm, power measurements (i.e., RSS estimates) are obtained at known locations in the offline phase. Then, in the online phase, power measurements are used together with the database to implement an NLS algorithm that takes NLOS effects into account by estimating real received powers from the database via the weighted $k$-NN {(or, the SVR)} algorithm. The proposed algorithm is compared with the NLS and fingerprinting algorithms to evaluate its localization performance. Although the fingerprinting and triangulation approaches have jointly been considered for VLP systems in \cite{Amorph_2016,JointFinTri}, the main novelty of the proposed approach can be stated as follows: (i) We form a database that contains NLOS information in received power measurements, and
%, which are modeled as scaled (by NLOS coefficients) and noise corrupted version of received powers due to LOS paths.
(ii) we design an NLS algorithm that utilizes %the LOS channel model, and
the database and current power measurements for position estimation.
%for estimating NLOS coefficients via a learning algorithm.}

%DA-NLS method combines the fingerprinting and NLS methods in a way that it utilizes a K-factor database which can incorporate the positive effects of having a database to more accurately model the NLOS channel in one sense, while allowing the exact computation of LOS powers at every possible point and hence eliminating the negative impacts of having a database in the other sense.

%%%%%%%%%%%%%%%%%%%%%%%%%%%%%%%%%%%%%%%%%%%%%%%%%%%%%%%%%%%%%%%%%%%%%%%%%%%%%%%%
\section{System Model and Problem Statement}\label{sec:System}
%\subsection{Optical Wireless Channel Characteristics}

{We consider an NLOS channel model that takes the LOS path and multipath components into account \cite{tashi,HarnesNLOS}.
%We consider an NLOS channel model that takes the LOS path and single reflections into account as in \cite{tashi}. Generally, the total power received at a PD due to reflections is significantly lower than that due to the LOS path. However, the effect of NLOS propagation cannot completely be omitted since some locations in an indoor space (e.g., corners) can have relatively high amounts of reflected lights. Therefore, it is important to consider an NLOS channel model and also sufficient to take only first reflections into account since first reflections and the direct path constitute a high majority of received energy \cite{HarnesNLOS}.
The channel DC gain for the LOS path} is given by \cite{tashi,LOSmodel}
\begin{equation} \label{eqn1}
    \hspace{-0.1cm}H_{LOS} =
    \begin{cases}
        \frac{(m+1)A}{2\pi D^2}\cos^m(\phi)T_s(\psi)g(\psi)\cos(\psi),&\hspace{-0.2cm}\text{if } 0\leq\psi\leq\psi_c\\
        0,              & \hspace{-0.2cm}\text{otherwise}
    \end{cases}
\end{equation}
where $D$ is the distance between a particular LED and a PD, $m$ is the Lambertian order (also called radiation lobe mode number), $A$ is the area of the PD, $\phi$ is the irradiance angle, $\psi$ is the incidence angle, $T_s(\psi)$ is the optical filter gain, $g(\psi)$ is the optical concentrator gain, and $\psi_c$ stands for the field-of-view (FOV) of the PD. The gain of the optical concentrator is calculated from $g(\psi)={n^2}/\sin^2\psi_c$ if $0\leq\psi\leq\psi_c$ and $g(\psi)=0$ otherwise,
%\begin{equation} \label{eqn2}
%    g(\psi) =
%    \begin{cases}
%        \frac{n^2}{\sin^2\psi_c}, & \text{if \,} 0\leq\psi\leq\psi_c\\
%        0,              & \text{otherwise}
%    \end{cases}
%\end{equation}
where $n$ is the refractive index \cite{tashi}.

The channel gain in an NLOS path involving a single reflection is expressed as follows \cite{tashi}:
\begin{equation}\label{eqn3}
    dH_{NLOS}=
    \begin{cases}
        \frac{(m+1)A}{2\pi^2 D_1^2 D_2^2}\cos^m(\phi)T_s(\psi)g(\psi)\,\rho\\
        \times\cos(\psi)\cos(\alpha)\cos(\beta) dA_{wall},&\hspace{-0.2cm}\text{if~} 0\leq\psi\leq\psi_c\\
        0, &\hspace{-0.2cm}\text{otherwise}
    \end{cases}
\end{equation}
where $\rho$ is the reflectance factor of the wall, $D_1$ is the distance between a particular LED and the reflective point on the wall, $D_2$ is the distance between the reflective point on the wall and the receiving PD, $\alpha$ is the irradiance angle to a reflective point on the wall, $\beta$ is the irradiance angle from the reflective point on the wall, and $dA_{wall}$ is a small reflective area on the wall.
%(please see Fig. \ref{fig:channel}).
%\begin{figure}[tbh!]
%\centering
%\includegraphics[scale = 0.4]{los- nlos channel fig 2}
%\caption{Model for an NLOS path.}
%\label{fig:channel}
%\vspace{-0.3cm}
%\end{figure}

By considering the LOS path and the {single} reflections, the received power at a PD due to transmission from an LED (call it LED~$i$) can be modeled as follows \cite{tashi}:
\begin{equation}\label{eqn4}
P_{RX,i} \approx P_{TX,i} \, H_{LOS,i} + \int_{walls}  P_{TX,i} \, dH_{NLOS,i} +\xi_i
\end{equation}
where %$N$ is the number of LEDs,
$P_{TX,i}$ denotes the transmitted power from LED~$i$ (assumed to be known), $H_{LOS,i}$ is the LOS channel gain between LED $i$ and the PD (see \eqref{eqn1}), $dH_{NLOS,i}$ is the channel gain between LED $i$ and the PD due a single reflection from a given area $dA_{wall}$ (see \eqref{eqn3}), and $\xi_i$ denotes the noise during the reception of the signal from LED $i$. The noise term is commonly modeled as a zero mean Gaussian random variable; i.e., $\xi_i \sim \mathcal{N}(0,\sigma^2_i)$ \cite{tashi,huang2,sinan3}.
%and it is commonly modeled as independent for different LEDs \cite{tashi,huang2,sinan3}.
%(NO NEEED) Also, the noise variance is calculated as \cite{tashi,huang2,sinan3}:
%\begin{equation} \label{eqn5}
%    \sigma^2_{total,i} = \sigma^2_{shot,i} + \sigma^2_{thermal,i} + %\sigma^2_{shadowing,i} + \sigma^2_{m. fading,i}.
%\end{equation}

{To provide a simpler and more generic expression, \eqref{eqn4} can be modified as
\begin{equation}\label{eq:RecPowSimp}
P_{RX,i} = \kappa_i \, P_{TX,i} \, H_{LOS,i} +\xi_i
\end{equation}
where $\kappa_i\geq 1$ specifies the combined effects of all LOS and NLOS components, including those with multiple reflections, as well. %\textcolor{magenta}{We call the parameters
%$\kappa_i$ \emph{NLOS coefficients}. \\ ---- Should we name them %NLOS coefficients anymore? ----}}

In the presence of $N$ LEDs with known positions $\boldsymbol{l}_1,\ldots\boldsymbol{l}_N$, the aim is to estimate the position $\boldsymbol{l}$ of the PD based on received power measurements related to transmissions from the LEDs (that is,  $P_{RX,1},\ldots,P_{RX,N}$), where $\boldsymbol{l}=(x,y,z)$ and $\boldsymbol{l}_i=(x_i,y_i,z_i)$ represent the three-dimensional positions of the PD and LED $i$, respectively. In some cases, it is possible to collect measurements at known positions in an environment of interest and form a database that can later be used for localization \cite{sinan3}. In the following section, we first {briefly mention} two common approaches in the literature, one using a database without {assuming a statistical model}, and the other using a {statistical} model without a database. Then, we propose a new localization approach that employs both a database and a {statistical} model with consideration of NLOS effects.

%%%%%%%%%%%%%%%%%%%%%%%%%%%%%%%%%%%%%%%%%%%%%%%%%%%%%%%%%%%%%%%%%%%%%%%%%%%%%%%%
\section{Positioning Algorithms}\label{sec:Algo}

\subsection{Fingerprinting Algorithm}

%In the fingerprinting algorithm, an offline training phase is performed first in order to form a database. Then, during the online localization phase, the unknown position is estimated based on the current measurements and the database \cite{sinan3}.

For employing the fingerprinting algorithm \cite{sinan3} for visible light positioning based on received power measurements, power levels are measured at some predetermined (known) locations of the PD and a database consisting of location vector and measurement vector pairs is formed during the training phase. In particular, the database can be represented as
\begin{equation} \label{eqn7}
    {\mathcal{D}} = \Big\{\big(\boldsymbol{l}^{(1)},\boldsymbol{P}^{(1)}_{RX}\big), \ldots, \big(\boldsymbol{l}^{(N_1)},\boldsymbol{P}^{(N_1)}_{RX}\big)\Big\}
\end{equation}
where $N_1$ is the number of predetermined locations for the PD, $\boldsymbol{l}^{(j)}$ is the $j$th predetermined location, and $\boldsymbol{P}^{(j)}_{RX}$ is the power measurement vector at $\boldsymbol{l}^{(j)}$ for $j=1,\ldots,N_1$. Here, $\boldsymbol{P}^{(j)}_{RX}=\big(P^{(j)}_{RX,1},\ldots,P^{(j)}_{RX,N}\big)$ with $P^{(j)}_{RX,i}$ denoting the received power at $\boldsymbol{l}^{(j)}$ due to the transmission from LED~$i$, where $i=1,\ldots,N$.\footnote{In practice, received power can be measured a number of times to mitigate the effects of noise via averaging.}

In the online phase, the received power measurements are collected (in real time) by the PD related to $N$ LEDs, which are denoted by $\boldsymbol{P}_{RX}=(P_{RX,1},\ldots,P_{RX,N}$), and the unknown location of the PD is estimated by utilizing the database {${\mathcal{D}}$}. Namely, a regression function is used along with the received power measurements to estimate the location. {The weighted $k$-NN regression and the SVR are commonly employed in the fingerprinting algorithm \cite{sinan3,FingerNN,MLVLP,LSSVM}.}

%========================
%In this manuscript, the weighted $k$-NN regressor is employed as a representative supervised learning approach \cite{sinan3,FingerNN,MLVLP} (please see Remark~2). In this approach, the Euclidean distances between the received power vector $\boldsymbol{P}_{RX}$ and the received power vectors in the database ${\mathcal{D}}_1$ (namely, $\boldsymbol{P}^{(1)}_{RX},\ldots,\boldsymbol{P}^{(N_1)}_{RX}$ in \eqref{eqn7}) are calculated to determine the closest $k$ vectors in the database. To that aim, set $\mathcal{S}_1$ is defined as
%\begin{align}\nonumber
%    {\mathcal{S}}_1=&\big\{i\in\{1,\ldots,N_1\}\,\big{|}\,\big\|\boldsymbol{P}_{RX}-\bolds%ymbol{P}^{(i)}_{RX}\big\| \leq \big\|\boldsymbol{P}_{RX}-\boldsymbol{P}^{(j)}_{RX}\big\| %\\&~~\forall i\in \mathcal{S}_1,~\forall %j\in\{1,\ldots,N_1\}\setminus\mathcal{S}_1,\,{\rm{and}}~|{\mathcal{S}_1}|=k\big\}.
%\end{align}
%Then, a weighted sum of $k$ locations corresponding to set $\mathcal{S}_1$ is calculated %as the location estimate as follows:
%\begin{equation}\label{eqn10}
%    \hat{\boldsymbol{l}} = %\frac
%    \small{\left(\sum_{m \in \mathcal{S}_1}\frac{\boldsymbol{l}^{(m)}}
%    {\|\boldsymbol{P}_{RX}-\boldsymbol{P}^{(m)}_{RX}\|}\right)}
%    \bigg{/}
%    \small{\left(\sum_{m \in \mathcal{S}_1}\frac{1}
%    {\|\boldsymbol{P}_{RX}-\boldsymbol{P}^{(m)}_{RX}\|}\right)}
%\end{equation}
%=====================================

\subsection{Nonlinear Least Squares (NLS) Algorithm}

In the absence of a database, the location of the PD can be estimated from the received power measurements, $\boldsymbol{P}_{RX}=(P_{RX,1},\ldots,P_{RX,N}$), {based on a statistical model. Considering independent zero-mean Gaussian noise components in \eqref{eq:RecPowSimp} and adopting the LOS channel model in \eqref{eqn1}} (i.e., $\kappa_i=1$ in \eqref{eq:RecPowSimp}), {the following NLS algorithm is employed in the literature \cite[eqn.~(46)]{sinanDirect}:
\begin{equation}\label{eq:NLS}
\hat{\boldsymbol{l}} =\underset{\boldsymbol{l}\in{\mathcal{L}}}{\arg\min}~
\sum_{i=1}^{N}\frac{\left({P}_{RX,i}-{P}_{TX,i}H_{LOS,i}(\boldsymbol{l})\right)^2}{\sigma_i^2}
\end{equation}
where ${\mathcal{L}}$ represents the set of all possible locations of the PD (e.g., all possible locations in a room), ${P}_{TX,i}$ is the transmit power of LED $i$, and $H_{LOS,i}(\boldsymbol{l})$ corresponds to the LOS channel gain in \eqref{eqn1}. (The argument $\boldsymbol{l}$ has been added to emphasize the dependence on the PD location.)}

\subsection{Proposed Algorithm: Database Assisted NLS}\label{sec:Proposed}

In this section, we propose an algorithm for visible light positioning based on ideas from both fingerprinting and NLS. The received power measurements {in \eqref{eq:RecPowSimp}} can be modeled as
\begin{equation}\label{eq:PrxNLS}
     \boldsymbol{P}_{RX}={\boldsymbol{P}}_{RX}^{(\rm{real})}(\boldsymbol{l})+\boldsymbol{\xi}
\end{equation}
where $\boldsymbol{\xi}$ is the vector of noise components and  ${\boldsymbol{P}}_{RX}^{(\rm{real})}(\boldsymbol{l})$ represents the real received power vector (without any modeling errors), which is a function of the PD location $\boldsymbol{l}$ \cite{sinan4,sinan5}. {Assuming that the noise components are independent zero-mean Gaussian random variables, specified as $\xi_i \sim \mathcal{N}(0,\sigma^2_i)$ for $i=1,\ldots,N$,} the maximum likelihood estimator (MLE) for the PD location based on $\boldsymbol{P}_{RX}$ in \eqref{eq:PrxNLS} can be obtained as
%\begin{align}
%\hat{\boldsymbol{l}} &= \underset{\boldsymbol{l}\in{\mathcal{L}}}{\arg\max}~p(\boldsymbol{P}_{RX}%\,|\,\boldsymbol{l})\\
%&=\underset{\boldsymbol{l}\in{\mathcal{L}}}{\arg\max}~
%\frac{(2\pi)^{-0.5N}}{\sigma_1\cdots\sigma_N}e^{-\sum_{i=1}^{N}\frac{\left({P}_{RX,i}-{P}_{RX,i}^%{(\rm{real})}(\boldsymbol{l})\right)^2}{2\sigma_i^2}}
%\end{align}
%which leads to
\begin{equation}\label{eq:MLest}
\hat{\boldsymbol{l}} =\underset{\boldsymbol{l}\in{\mathcal{L}}}{\arg\min}~
\sum_{i=1}^{N}\frac{1}{\sigma_i^2}{\left({P}_{RX,i}-{P}_{RX,i}^{(\rm{real})}(\boldsymbol{l})\right)^2}
\end{equation}
with ${P}_{RX,i}^{(\rm{real})}$ denoting the $i$th component of ${\boldsymbol{P}}_{RX}^{(\rm{real})}(\boldsymbol{l})$ and ${\mathcal{L}}$ representing the set of all possible locations of the PD.

In practice, finding an accurate and tractable mathematical model for  ${\boldsymbol{P}}_{RX}^{(\rm{real})}(\boldsymbol{l})$ in \eqref{eq:MLest} is very challenging {due to NLOS effects}. Even the consideration of only the LOS component and the single reflection components makes the expression very complicated (cf.~\eqref{eqn4}). Therefore, {{we aim to utilize a database to learn ${\boldsymbol{P}}_{RX}^{(\rm{real})}(\boldsymbol{l})$ in \eqref{eq:MLest}.}}

In the first (offline) phase of the proposed algorithm, power measurements are collected at some predetermined locations in the environment {{as in \eqref{eqn7} of}} the fingerprinting algorithm.

In the second (online) phase of the proposed algorithm, the aim is to estimate the unknown location $\boldsymbol{l}$ of the PD based on the received power measurements, $\boldsymbol{P}_{RX}=(P_{RX,1},\ldots,P_{RX,N})$, related to the transmissions from the LEDs. These power measurements are used in an NLS algorithm that {effectively} takes NLOS effects into account based on the information in the database ${\mathcal{D}}$ in \eqref{eqn7}. In developing the proposed approach, {the database is utilized to obtain an estimate for the value of ${\boldsymbol{P}}_{RX}^{(\rm{real})}(\boldsymbol{l})$ in \eqref{eq:MLest} for each given value of $\boldsymbol{l}$.} To that aim, we use the $k$-NN algorithm over set ${\mathcal{D}}$ as follows:\footnote{{Alternatively, a similar learning approach such as SVR can also be used.}}
\begin{equation}\label{eq:kNN_kappa}
{ {\widehat{\boldsymbol{P}}}_{RX}^{(\rm{real})}(\boldsymbol{l})}=
    \small{\left({\sum_{m \in \mathcal{S}({\boldsymbol{l}})}\frac{{\boldsymbol{P}^{(m)}_{RX}}}
    {\|\boldsymbol{l}-\boldsymbol{l}^{(m)}\|}}\right)}
    \bigg{/}
    \small{\left({\sum_{m \in \mathcal{S}({\boldsymbol{l}})}\frac{1}
    {\|\boldsymbol{l}-\boldsymbol{l}^{(m)}\|}}\right)}
\end{equation}
where {${\widehat{\boldsymbol{P}}}_{RX}^{(\rm{real})}(\boldsymbol{l})=\big({\widehat{{P}}}_{RX,1}^{(\rm{real})}(\boldsymbol{l}),\ldots,{\widehat{{P}}}_{RX,N}^{(\rm{real})}(\boldsymbol{l})\big)$} and
\begin{align}\label{eq:S2}
    &{\mathcal{S}}({\boldsymbol{l}})=\big\{i\in\{1,\ldots,N_1\}\,\big{|}\,\big\|{\boldsymbol{l}}-{\boldsymbol{l}}^{(i)}\big\| \leq \big\|{\boldsymbol{l}}-{\boldsymbol{l}}^{(j)}\big\| \\\nonumber
    &~~\forall i\in \mathcal{S}({\boldsymbol{l}}),~\forall j\in\{1,\ldots,N_1\}\setminus\mathcal{S}({\boldsymbol{l}}),\,{\rm{and}}~|{\mathcal{S}}({\boldsymbol{l}})|=k\big\}\,.
\end{align}
Then, the {real received power} estimates in \eqref{eq:kNN_kappa} are used for the real received power values in \eqref{eq:MLest}, and the proposed NLS algorithm is stated as
\begin{equation}\label{eq:NLSpropFinal}
\hat{\boldsymbol{l}} =\underset{\boldsymbol{l}\in{\mathcal{L}}}{\arg\min}\,
\sum_{i=1}^{N}\frac{1}{\sigma_i^2}{\left({P}_{RX,i}-{{\widehat{{P}}}_{RX,i}^{(\rm{real})}(\boldsymbol{l})}\right)^2}
\end{equation}
where {${\widehat{{P}}}_{RX,i}^{(\rm{real})}(\boldsymbol{l})$} is the $i$th element of {${\widehat{\boldsymbol{P}}}_{RX}^{(\rm{real})}(\boldsymbol{l})$} in \eqref{eq:kNN_kappa}.

The proposed algorithm is summarized in Algorithm~1, which is named \emph{database assisted NLS} (DA-NLS) as it employs an NLS algorithm by utilizing a database for NLOS consideration purposes.

\begin{algorithm}[H]
\small
%\SetAlgoLined
\KwResult{Location estimate $\hat{\boldsymbol{l}}$}
%\BlankLine
\emph{- Training (Offline) Phase}\;
%\BlankLine

Set $N_1$ locations, $\boldsymbol{l}^{(1)},\ldots,\boldsymbol{l}^{(N_1)}$, for database $\mathcal{D}$
%\BlankLine

Initialize $\mathcal{D}=\emptyset$

\For{$j=1,\dots,N_1$}{
    \For{$i=1,\dots,N$}
    {
        Get measurement $P_{RX,i}^{(j)}$ for location $\boldsymbol{l}^{(j)}$ and LED $i$ %from PD at location $\boldsymbol{l}^{(j)}$
        %\BlankLine

        %Compute $H_{LOS,i}^{(j)}$ via \eqref{eqn1} %and $\boldsymbol{l}^{(j)}$
        %\BlankLine

        %Calculate $\hat{\kappa}^{(j)}_i$ as %$\hat{\kappa}^{(j)}_i=P_{RX,i}^{(j)}/(P_{TX,i}H_{LOS,i}^{(j)})$
    }
Add $\big{(}\boldsymbol{l}^{(j)}$, ${\boldsymbol{P}}_{RX}^{(j)}\big{)}$ to $\mathcal{D}$
}
%\BlankLine
\emph{- Localization (Online) Phase}\;
%\BlankLine

Obtain $\boldsymbol{P}_{RX}=({P}_{RX,1},\ldots,{P}_{RX,N})$ from PD to be located
%\BlankLine

Run an optimization algorithm (e.g., PSO \cite{PSO1}) to solve \eqref{eq:NLSpropFinal} by evaluating $\widehat{\boldsymbol{P}}_{RX}^{(\rm{real})}(\boldsymbol{l})$ from \eqref{eq:kNN_kappa} and \eqref{eq:S2}.

%Set $E_{\min}=\infty$

%\For{$\boldsymbol{l}\in{\mathcal{L}}$}{
    %\BlankLine

        %Obtain set $\mathcal{S}_2$ in \eqref{eq:S2} from database $\mathcal{D}_2$

    %Compute $\hat{\boldsymbol{\kappa}}(\boldsymbol{l})$ in \eqref{eq:kNN_kappa}

    %\BlankLine

    %Compute $H_{LOS,i}(\boldsymbol{l})$ via \eqref{eqn1} for $i \in \{1,\ldots,N\}$

    %\BlankLine

    %Compute $E(\boldsymbol{l})=\sum_{i=1}^{N}\frac{\left({P}_{RX,i}-\hat{\kappa}_i(\boldsymbol{l}) \, P_{TX,i} \, H_{LOS,i}(\boldsymbol{l})\right)^2}{\sigma_i^2}$

    %If $E(\boldsymbol{l})<E_{\min}$, then $\hat{\boldsymbol{l}}=\boldsymbol{l}$ and $E_{\min}=E(\boldsymbol{l})$
%}

%\BlankLine

 %\boldsymbol{l} minimizing sum($\frac{(P_{RX,i} -  \hat{\kappa}_i(\boldsymbol{l})P_{TX,i}H_{LOS,i}(\boldsymbol{l}))^2} {\sigma^2_i}$) is $\hat{\boldsymbol{l}}$

%Obtain $\hat{\boldsymbol{l}}$ from \eqref{eq:NLSpropFinal}

%\BlankLine

 \caption{Database Assisted NLS (DA-NLS)}
\end{algorithm}

\textit{Remark~1:} The proposed DA-NLS algorithm utilizes the {knowledge of the statistical model for the noise components in the generic received power formula in \eqref{eq:PrxNLS}. As long as the noise components are zero-mean and independent Gaussian random variables with known variances,} the proposed algorithm can learn the {real received power expression} from the database and obtain an accurate model that leads to accurate location estimation. However, if there exist mismatches between the employed {statistical} model and the actual one, location estimation accuracy of the DA-NLS algorithm degrades and, for significant mismatches, can even be worse than that of the fingerprinting algorithm, which does not use a {statistical} model during location estimation. %However, in practical systems, validity of the Lambertian channel model in \eqref{eqn1} and methods for accurately estimating the parameters of that model are shown \cite{haas}.
{In addition, it is noted from \eqref{eq:PrxNLS} that the DA-NLS algorithm is a generic approach which is not specific to VLP systems.}

%On the other hand, the NLS algorithm in \eqref{eq:NLS} is specifically designed for LOS channels.

\textit{Remark~2:} If channel conditions change, a new database should be obtained for the fingerprinting algorithm and the proposed DA-NLS algorithm.

%%%%%%%%%%%%%%%%%%%%%%%%%%%%%%%%%%%%%%%%%%%%%%%%%%%%%%%%%%%%%%%%%%%%%%%%%%%%%%%
\section{Simulation Results and Conclusions} \label{sec:conc}

%{Fingerprinting icin k-NN'de Euclidean distance ile ters orantili kestirim yapildigini yazalim, (10) gibi.}

In this section, simulations are performed to evaluate the localization algorithms in the previous section based on the channel model specified by \eqref{eqn1}-\eqref{eqn4}. A room with dimensions $5\times5\times3$ meters is considered, where $3$ meters corresponds to the height \cite{tashi}. There are $4$ LEDs placed on the ceiling, which  point downwards. The locations of the LEDs are $(-1.25,-1.25, 1.5)$, $(-1.25,1.25,1.5)$, $(1.25,-1.25,1.5)$ and $(1.25,1.25,1.5)$ meters, with $(0,0,0)$ corresponding to the geometric center of the room. The PD to be localized points upwards and is at a fixed height of $0.85\,$m, which is a known parameter \cite{tashi,sinan5}. Therefore, a two-dimensional localization scenario is considered (e.g., the PD is on the top of a robot) \cite{sinan5}. For the fingerprinting and the DA-NLS algorithms, two separate databases are collected over two different uniform grids of sizes $10\times10$ and $28\times28$ (over the $5\times5$ meters area at the PD height). The variances of the noise components are taken to be equal; that is, $\sigma_i^2=\sigma^2=9.15\times10^{-7} \,W $ for $i=1,\ldots,N$ \cite{parametreler}. The other system parameters in Section~\ref{sec:System} are specified as follows \cite{tashi,parametreler}: $\psi_c={70}$ deg., $A=1\,{\rm{cm}}^2$, $n=1.5$, $T_s(\psi)=1$, $m=0.646$, and $\rho=0.8$. In addition, all the LEDs transmit the same amount of power in each simulation scenario; i.e., $P_{TX,i} = P_{TX}$ for $i=1,\ldots,N $.
%\begin{center}
% \begin{tabular}{||c c||}
% \hline
% Parameters &  Values\\ [0.5ex]
% \hline\hline
% Transmitted Power Per LED ($P_{TX,i}$) & 20 mW \\
% \hline
% {Semi-Angle} & {70 deg.}\\
% \hline
% FOV at Receiver ($\psi_c$) & 179 deg.\\
% \hline
% Area of PD ($A$) & 1 ${\rm{cm}}^2$ \\
% \hline
% Refractive Index ($n$) & 1.5 \\
% \hline
% {Optical Filter Gain ($T_s(\psi)$)} & {1} \\
% \hline
% {Radiation Lobe Mode Number ($m$)} & {0.646} \\
% \hline
% reflectance factor of Walls ($\rho$) & 0.8 \\ [1ex]
% \hline
%\end{tabular}
%\end{center}
To evaluate performance of the localization algorithms, $1000$ random points are generated in MATLAB R2018b for the locations of the PD (the \emph{rng default} command is used). For solving \eqref{eq:NLS} in the NLS algorithm and \eqref{eq:NLSpropFinal} in the DA-NLS algorithm, the particle swarm optimization (PSO) algorithm in MATLAB is utilized. {Also, the fingerprinting algorithm employs a weighted $k$-NN regressor (as in \eqref{eq:kNN_kappa}), which assigns weights inversely proportional to Euclidean distances between power entries in the database and measured powers to $k$ corresponding database locations.} In addition, each received power measurement is repeated $1000$ times for reducing the effects of noise (please see Footnote~1).

%\subsection{Simulation Results}

\begin{figure}%[tbh!]
%\vspace{-0.5cm}
\centering
\includegraphics[scale = 0.52]{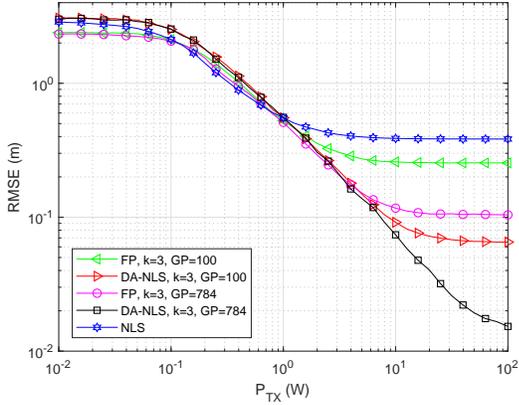}
\caption{{RMSE versus $P_{TX}$ for different algorithms.}}
\label{fig:rmse}
%\vspace{-0.5cm}
\end{figure}

In Fig.~\ref{fig:rmse}, the root mean-squared error (RMSE) values are plotted versus $P_{TX}$ for the fingerprinting (FP), DA-NLS and NLS algorithms by considering two different databases. %({$P_{TX}$ is defined as the light power emitted by a single LED in units of Watt}.)
In the figure, ${\rm{GP}}=100$ and ${\rm{GP}}=784$ correspond to the grids of $10\times10$ and $28\times28$, described previously. In the fingerprinting algorithm, for each of the grids, the optimal $k$ value for $P_{TX} = 20\,$W is used in the $k$-NN algorithm for all $P_{TX}$ values (see Fig.~\ref{fig:k}). For the DA-NLS algorithm, the same $k$ values ($k=3$) are employed as in the fingerprinting algorithm. As can be observed from Fig.~\ref{fig:rmse}, the proposed DA-NLS algorithm achieves significantly lower RMSEs than the NLS and the fingerprinting algorithms {for high transmit powers}. {For {$P_{TX} < 1\,$W}, the RMSEs are high (above $0.5$ meter) and the algorithms achieve close performance in general, with the DA-NLS algorithm having slightly higher RMSE values. In the case of very low transmit powers ($P_{TX} < 0.1\,$W), the fingerprinting algorithm achieves the lowest RMSEs as it does not assume any channel model.} Also, as expected, the estimation accuracy of the fingerprinting and the DA-NLS algorithms improves as the grid density increases. It is noted that the DA-NLS algorithm outperforms all the other algorithms even by using the smaller database when {$P_{TX} > 5\,$W}.
%For {$P_{TX} < 5\,$W}, the fingerprinting algorithm with the larger database slightly outperforms the other algorithms in general.

%Another observation is that the performances of Fingerprinting and NLS algorithms are significantly less sensitive to the changes in the SNR for $SNR \geq 5$ compared to DA-NLS algorithm. It is deduced from the Fig. \ref{fig:rmse} that the range of RMSE for $SNR \geq 5$ for Fingerprinting and NLS algorithms is {(I considered FP w/ both GP and NLS to compute a single range)} 0.0317 m whereas the range of RMSE for DA-NLS algorithm with GP parameter set to 100 is 0.0801 m and with GP parameter set to 784 is 0.1139 m for $SNR \geq 5$.

\begin{figure}%[tbh!]
%\vspace{-0.5cm}
\centering
\includegraphics[scale = 0.52]{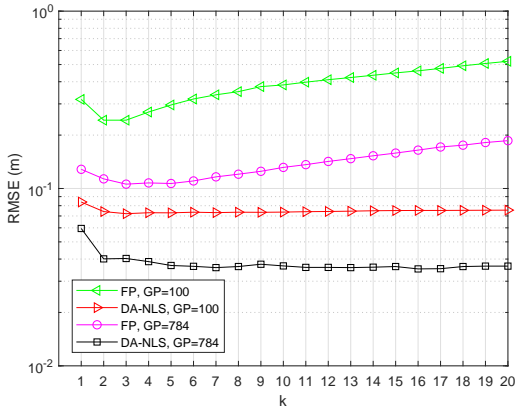}
\caption{RMSE versus $k$ for fingerprinting and DA-NLS algorithms at {$P_{TX} = 20\,$W}.}
\label{fig:k}
%\vspace{-0.5cm}
\end{figure}

In Fig.~\ref{fig:k}, the RMSE is plotted versus $k$ for the fingerprinting and the DA-NLS algorithms at {$P_{TX} = 20\,$W} considering the two databases. It is observed that for $k \geq 2$, the DA-NLS algorithm yields a steady RMSE performance implying that any $k \geq 2$ can comfortably be used in the DA-NLS algorithm. However, the accuracy of the fingerprinting algorithm degrades as $k$ increases for $k > 3$ and $k>5$ for the small and large databases, respectively. {Since the fingerprinting algorithm generates its location estimate based on the $k$-NN approach directly, it is more sensitive to $k$.} In addition, the DA-NLS algorithm outperforms the fingerprinting algorithm for both databases and for all values of $k$ at $P_{TX} = 20\,$W. %In addition, for both algorithms and for all k, performance increases as number of GP increases.

%\begin{figure}[tbh!]
%\centering
%\includegraphics[scale = 0.38]{CDF}
%\caption{CDFs for different algorithms at {$P_{TX} = 5$ and $20$ W}.}
%\label{fig:CDF}
%\end{figure}
%The cumulative distribution functions (CDFs) of absolute errors are presented in Fig.~\ref{fig:CDF} for all  three algorithms at {$P_{TX} = 5$ and $20\,$W}, where $k=4$ and the larger database is used for the fingerprinting and the DA-NLS algorithms. It is observed from Fig.~\ref{fig:CDF} that at {$P_{TX} = 20\,$W}, the absolute errors are lower than ${0.0505}$, ${0.1255}$, and ${0.575}$ meters with $90\%$ probability for the DA-NLS, fingerprinting, and NLS algorithms, respectively. Hence, the DA-NLS algorithm provides {significantly} more accurate estimates than the other two algorithms at {$P_{TX} = 20\,$W}. On the other hand, at {$P_{TX} = 5\,$W}, both the estimation accuracy and the differences between the algorithms decreases. In this scenario, the absolute errors corresponding to the $90\%$ level are {$0.196$, $0.2098$, and $0.5965$} meters for the DA-NLS, fingerprinting, and NLS algorithms, respectively.
%These observations are in compliance with the results in Fig.~\ref{fig:rmse}, which demonstrate that the estimation performances of the DA-NLS and the fingerprinting algorithms get close as {$P_{TX}$} decreases.

{Next}, the effects of the reflectance factor ($\rho$) on the localization performance are examined by plotting the RMSE against $\rho$ in Fig.~\ref{fig:rho} for two different {$P_{TX}$ values}, where $k=4$ and the larger database is used for the fingerprinting and the DA-NLS algorithms. The figure reveals that the variation of the $\rho$ parameter does not have significant effects on the accuracy of the DA-NLS algorithm for both {$P_{TX}$ values}; i.e., the DA-NLS algorithm is robust against the NLOS effects. Similarly, the fingerprinting algorithm is robust against $\rho$ at {$P_{TX} = 5\,$W}. However, at {$P_{TX} = 20\,$W}, there is an increase in the RMSE of the fingerprinting algorithm as $\rho$ increases. The NLS algorithm is very sensitive to $\rho$ since it assumes an LOS channel model (see \eqref{eq:NLS}) and the NLOS effects increase with $\rho$. The sensitivity to $\rho$ increases at higher {transmit powers} since the channel related errors become more significant than the noise related errors in that case.
%For Fingerprinting, the range of RMSE is 0.024 m for SNR = 0 whereas it is 0.055 m for SNR = 15. Similarly, for NLS, the range of RMSE is 0.243 m for SNR = 0 whereas it is 0.327 m for SNR = 15.
%Another important observation is that the RMSE performance of NLS algorithm is heavily sensitive to changes in the $\rho$ parameter (which is even more effective at high SNR value of 15) and as $\rho$ parameter increases, the accuracy seriously degrades.
Moreover, if the $\rho$ parameter is zero, which corresponds to an LOS scenario (i.e., no NLOS propagation), the NLS algorithm achieves the best performance since it becomes the exact MLE in that case. % (i.e., \eqref{eq:MLest} and \eqref{eq:NLS} become equivalent).
However, as the NLOS effect ($\rho$) increases, its performance degrades quickly.
%the performance of NLOS algorithm can even surpass that of DA-NLS algorithm for high SNR values. This is reasonable since NLS algorithm can compute the exact LOS powers at every point in the room via the LOS channel model presented in equations (\ref{eqn1}), (\ref{eqn2}) and (\ref{eqn3}) whereas DA-NLS algorithm needs to compute the LOS powers via several database points. Given that there is only a small amount of noise presence (i.e. SNR = 15), then the performance of NLS algorithm surpasses that of DA-NLS algorithm for $\rho = 0$ and $\rho = 0.05$ as shown in Fig. \ref{fig:rho}.

\begin{figure}%[tbh!]
%\vspace{-0.4cm}
\centering
\includegraphics[scale = 0.52]{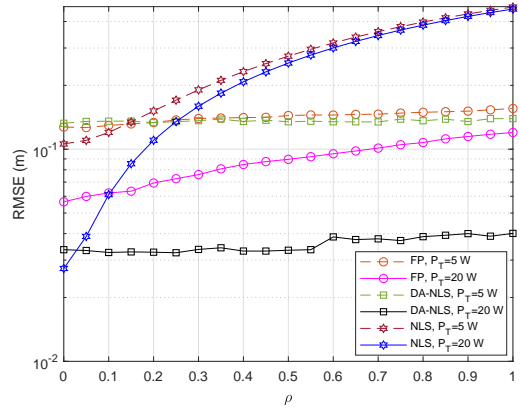}
\caption{RMSE versus $\rho$ for different algorithms at {$P_{TX} = 5$ and $20\,$W}.}
\label{fig:rho}
%\vspace{-0.5cm}
\end{figure}

{To further evaluate the performance of the proposed algorithm, the $k$-NN regressor is replaced with a machine learning based regression technique, namely SVR, during the estimation of location and $\widehat{\boldsymbol{P}}_{RX}^{(\rm{real})}(\boldsymbol{l})$ at the online phases of the FP and DA-NLS algorithms, respectively. For implementing the SVR algorithm, MATLAB's \textit{fitrsvm} and \textit{predict} methods are employed. The kernel function is set to Gaussian and the \textit{OptimizeHyperparameters} option is set to `auto', the latter of which is used to optimize hyperparameters of the SVR. In order to enhance optimization performance, the number of iterations is increased to $100$ from the default value of $30$. Also, to ensure reproducibility, the \textit{rng default} command is used just before training the model and the acquisition function of the model is set to `expected-improvement-plus'. Note that for each different $P_{TX}$, the database from which SVRs are trained changes. Therefore, a new SVR model is trained for each $P_{TX}$ so as to achieve enhanced prediction performance. In the DA-NLS algorithm, since the total received power at any point in the room is the sum of received powers from individual LEDs, a separate SVR model is trained for each LED. For the FP algorithm, one SVR is trained for each coordinate to be estimated. The results of the simulations are shown in Fig.~\ref{fig:svr}. From comparison of this figure with Fig.~\ref{fig:rmse}, it is observed that the prediction performance of the FP algorithm increases for both databases at high transmit powers (i.e., $P_{TX} > 25W$) with respect to the performance of the $k$-NN based estimators. On the other hand, the RMSEs of the DA-NLS algorithm increases for the smaller database and slightly decreases for the larger database for high transmit powers (i.e., $P_{TX} > 25W$) when compared to the $k$-NN based estimators. Still, the NLS has highest RMSEs for $P_{TX} > 1W$. Overall, using SVR instead of $k$-NN does not lead to significant improvements in the performance of the DA-NLS algorithm, implying that the DA-NLS algorithm has desirable localization performance even when the simple $k$-NN algorithm is employed to estimate received powers.}

\begin{figure}%[tbh!]
%\vspace{-0.2cm}
\centering
\includegraphics[scale = 0.52]{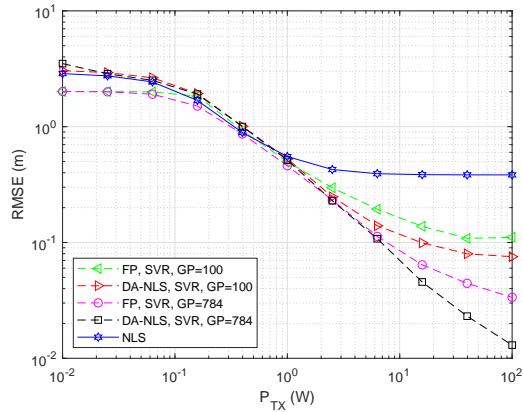}
\caption{{RMSE versus $P_{TX}$ when the SVR algorithm is used.}}
\label{fig:svr}
%\vspace{-0.5cm}
\end{figure}

%Overall, the proposed DA-NLS algorithm utilizes ideas from the NLS and fingerprinting algorithms, and provides a robust localization performance for VLP systems in NLOS scenarios.

%%%%%%%%%%%%%%%%%%%%%%%%%%%%%%%%%%%%%%%%%%%%%%%%%%%%%%%%%%%%%%%%%%%%%%%%%%%%%%%%
%\section{Conclusion}

%%%%%%%%%%%%%%%%%%%%%%%%%%%%%%%%%%%%%%%%%%%%%%%%%%%%%%%%%%%%%%%%%%%%%%%%%%%%%%%%

\bibliographystyle{ieeetr}
\bibliography{sample}

%\section{Acknowledgements}
%\noindent Acknowledgements will be here.

%\clearpage

\end{document}